# Deep learning empowered synthetic dimension dynamics: morphing of light into topological modes

Shiqi Xia[1+], Sihong Lei[1+], Daohong Song[1,2], Luigi Di Lauro[3], Imtiaz Alamgir[3], Liqin Tang[1,2], Jingjun Xu[1], Roberto Morandotti[3], Hrvoje Buljan[1,4*], Zhigang Chen[1,2*]

*1 The MOE Key Laboratory of Weak-Light Nonlinear Photonics, TEDA Institute of Applied Physics and School of Physics, Nankai University, Tianjin 300457, China*
*2 Collaborative Innovation Center of Extreme Optics, Shanxi University, Taiyuan, Shanxi 030006, China*
*3 INRS-EMT, 1650 Blvd. Lionel-Boulet, Varennes, Quebec J3X 1S2, Canada*
*4 Department of Physics, Faculty of Science, University of Zagreb, Bijenička c. 32, Zagreb 10000, Croatia*
[+]*These authors contributed equally to this work*
*\* hbuljan.phy@pmf.hr, zgchen@nankai.edu.cn*

**Abstract:** Synthetic dimensions (SDs) opened the door for exploring previously inaccessible phenomena in high-dimensional synthetic space. However, construction of synthetic lattices with desired coupling properties is a challenging and unintuitive task, largely limiting the exploration and current application of SD dynamics. Here, we overcome this challenge by using deep learning artificial neural networks (ANNs) to validly design the dynamics in SDs. We use ANNs to construct a lattice in real space that has a predesigned spectrum of mode eigenvalues. By employing judiciously chosen perturbations (wiggling of waveguides), we show experimentally and theoretically resonant mode coupling and tailored dynamics in SDs, which leads to effective transport or confinement of a complex beam profile. As an enlightening example, we demonstrate morphing of light into a topologically protected edge mode in ANN-designed Su-Schrieffer-Heeger photonic lattices. Such ANN-assisted construction of SDs advances towards utopian networks, opening new avenues in fundamental research beyond geometric limitations. Our findings may offer a flexible and efficient solution for mode lasing, optical switching, and communication technologies.

Synthetic dimensions (SDs) are drawing a great deal of interest in topological photonics and other branches of physics for exploiting fundamental phenomena in high dimensional spaces [1, 2]. Several theoretical proposals have been put forward for the study and implementation of synthetic gauge fields, quantum Hall physics, discrete solitons, Weyl semimetals and topological phase transitions in four dimensions and beyond, predicting rich physics inaccessible in a conventional three-dimensional real space [3-8]. In particular, it is quite challenging to realize a complex structure or network of resonators with anisotropic, long-range or dissipative couplings in a real space, but that is particularly amenable using SDs. Exemplary successes include experimental demonstrations of non-Hermitian topological winding [9, 10], skin effects [11], parity-time symmetry [12, 13], topological "triple" phase transition [14] and nontrivial topology arising from a system's dissipation [15]. Thus far, SDs have been constructed using a variety of parameters or degrees of freedom in a given system, such as frequency modes, spatial modes, orbital angular momenta, and time-delayed pulses [11, 14, 16-21], along with many proposed applications in for example optical communications and topological insulator lasers [22, 23].

One highly desirable goal of these studies is to construct a utopian network of resonators or coupled modes, where any pair of modes could be coupled in a controlled fashion [24]. However, the possibilities of coupling are in most systems depend on their natural properties. For example, the realization of resonant coupling strongly depends on a given spectrum of eigenvalues. In one aspect of our work, we manipulate the spectra of modes by using artificial neural networks (ANNs) as one step towards the utopian networks. Moreover, mode manipulation has emerged as an active research subject in many photonic systems, as it brings about new possibilities for improving the design and functionality of devices. For instance, appropriate mode manipulation can increase the capacities of data transmission [25, 26], enhance the efficiency of energy harvesting [11], and enlarge the radiance of laser arrays [27]. Recently, mode manipulation using the concepts of non-Hermiticity, nonlinear topological photonics, as well as optical thermalization has been proposed and demonstrated [28-31]. Despite this progress, the resonant excitation between two specifically chosen modes is often hard to control, and the number of modes that can be manipulated synergetically is quite limited.

In this work, we implement on-demand waveguide arrays with desired features in SDs. We use a pretrained ANNs to design a waveguide array (i.e., a photonic lattice) in real space, which has a predetermined and desired spectrum of eigenvalues (see Fig. 1). This enables design and control of the mode coupling in SDs by employing perturbations with a specific frequency (or frequencies) in correspondence to spacing between mode eigenvalues. Thus, we achieve mode manipulation in a linear system, without the need to introduce gain/loss, nonreciprocal coupling, or nonlinearity. We experimentally implement two mode arrays forming an SD: one with uniform synthetic mode coupling strengths, and the other with edge defect weakly coupled to the bulk in the SD, showing correspondingly light transport and confinement in the SD. Finally, we utilize ANNs to design a modified type of Su-Schrieffer-Heeger (SSH) topological lattice [32] which has a linear dispersion in bulk bands, and show controlled coupling in the SD and thereby morphing of light from a given bulk mode into a topological edge state.

The eigenmodes and eigenvalues of any Hamiltonian $H$ can in principle be found by diagonalization, which produces a transformation matrix $\Phi$ comprising eigenmodes of $H$. The diagonal matrix $E = \Phi^\dagger H \Phi$ contains the eigenvalues, conveniently written as an array $B = \text{diag}(E)$. The inverse problem is far more complicated: given a preassigned eigenvalue array $B = \text{diag}(E)$, what is the Hamiltonian $H$ with such eigenvalues? Moreover, this task is here constrained with physical reality: we demand that only the coupling between the nearest neighbor waveguides is present in the Hamiltonian, which is consistent with experimental conditions employing evanescently coupled waveguides. In this work, we employ deep learning [33, 34] to find the Hamiltonian which yields a desired eigenvalue array $B$.

The deep learning method, a subset of machine learning that uses ANNs, has been applied in many physical systems, including photonics, image recognition, data analysis and inverse design [35-38]. A general description of the ANN paradigm we used can be found in the Supplementary Materials (SM) [39]. Here, we use a preassigned eigenvalue array $B$ as the input layer data of the ANNs (Fig. 2a1). The number of modes is chosen to be $N = 8$, but our method can be readily applied for larger systems [39]. The couplings between nearest neighbor sites, which define the Hamiltonian $H$, are assigned as the output layer. To determine the values of weights in all hidden layers, 800 sets of eigenvalue-Hamiltonian data (obtained numerically with the tight-binding model) are sent to the ANNs to train the networks [39].

When the ANN training is completed, two preassigned eigenvalue arrays $B$ illustrated in Fig. 2(a1, b1) are sent back to the ANNs. The first array (Fig. 2(a1)) is an equally spaced array, whereas the second one (Fig. 2(b1)) has two outlying eigenvalues at the edges of the SD [40]. For each of the

eigenvalue arrays, the ANNs (Fig. 2a1) yield a corresponding tight-binding Hamiltonian of the form

$$H_A = \sum_{n=1}^{N-1} t_n c_{n+1}^\dagger c_n + H.c. \qquad (1)$$

where $c_n^\dagger$ and $c_n$ are the creation and annihilation operators on the $n$-th site in real space, and $H.c.$ stands for the Hermitian conjugation. The eigenvalues of $H_A$ are given by $E_A = \Phi_A^\dagger H_A \Phi_A$. To evaluate the effectiveness of our ANN method, we use the following figure of merit:

$$\gamma_B = \frac{1}{N} \sum_{i=1}^{N-1} |\Delta\beta_{Ai} - \Delta\beta_i|/\Delta\beta_i \qquad (2)$$

where $\beta_i$ is the propagation constant of the $i$-th mode, and $\Delta\beta_i$ ($\Delta\beta_{Ai}$) is the eigenvalue difference between the $i$-th and $(i+1)$-th mode in the preassigned (ANN-calculated) eigenvalue array. The figure of merit compares the difference between the eigenvalues because it determines the coupling of modes in SD driven by perturbations. We find that $\gamma_B$ is approximately $0.90\%$ for the mode array with equal spacing, and $\gamma_B = 3.15\%$ for the array with outlying edges. The low deviations calculated for these two different mode arrays show the high effectiveness of our ANN method.

By appropriately wiggling the waveguides along the propagation direction (Fig. 1), we engineer the coupling between eigenmodes of the Hamiltonian $H_A$, that is, we engineer the dynamics in SD [21, 39]. Let $D_n$ denote the equilibrium distance between the first and $n$th waveguide. If the center of each waveguide wiggles along the propagation axis as $R\sin(\Omega z + \theta)$ in $(x, z)$ plane, this is equivalent to introducing an oscillating force of magnitude $F(z) = -\partial V(x, z)/\partial x$, which can be derived from a scalar potential $V(x, z) = V_0 \sin(\Omega z + \theta)(x - D_N/2)$. Here, we have for convenience set the zero point of the potential $V(D_N/2) = 0$ to be the central point of our lattice denoted with $D_N/2$. The $D_N$ is the distance between the first and last ($N$-th) waveguides. The amplitude $V_0 = k_0 \Omega^2 R$ can be calculated for our system straightforwardly in a vector potential gauge [21, 39]; here $k_0 = 2\pi n_0/\lambda$ is the wavenumber ($n_0$ is the refractive index of the medium, and $\lambda$ is the wavelength). After applying the wiggling, the Hamiltonian becomes

$$H_w(z) = H_A + H_1(z), \quad H_1(z) = \sum_{n=1}^{N} (D_n - D_N/2) k_0 \Omega^2 R \sin(\Omega z + \theta) c_n^\dagger c_n + H.c. \qquad (3)$$

Here, $z$ indicates the propagation distance along the photonic lattices, which is equivalent to the role of *time* in quantum mechanics. In the language of quantum mechanics, we utilize appropriately tailored time-dependent perturbations to transfer the energy between eigenmodes in a controllable fashion, which is enabled by the preassigned energies that enter the ANNs. The on-site terms in SD Hamiltonian are given by the diagonal matrix $\Phi_A^\dagger H_A \Phi_A$ (Fig. 2a2, b2), and the wiggling induced coupling coefficients are contained in the matrix $\Phi_A^\dagger H_1(z) \Phi_A$ (Fig. 2a3, b3) [21]; note that the oscillating term $\sin(\Omega z + \theta)$ can be extracted from the latter expression.

For the mode array with equal spacing between the propagation constants, the two nonzero off-diagonals indicate dominance of the nearest neighbor coupling in SD (Fig. 2a3). The wiggling frequency is identical to the equal spacing between the propagation constants. By using the proper rotating-wave approximation, which is here equivalent to interaction picture in quantum mechanics, the linear on-site potential in $\Phi_A^\dagger H_0 \Phi_A$ is eliminated (dashed line in Fig. 2a2) [21, 39]. Therefore, the structure in Fig. 2a can be regarded as a one-dimensional waveguide array in SD. If the mode circled in Fig. 2a4 is initially excited, it undergoes discrete diffraction in SD, shown in the numerical simulation.

For the mode array with outlying edges sketched in Fig. 2b, in the interaction picture, the on-site energies of the bulk modes in SD are zero, but the on-site energies of the outlying modes differ (dashed line in Fig. 2b2). The modes inside the blue square in Fig. 2b3 are mutually well coupled with nearest neighbor interactions (Fig. 2b3). However, the two outlying modes are only weakly coupled to the bulk. This is the consequence of the ANN engineered spectra of the propagation constants. We find some minor long-range couplings between modes, however the percentage of the maximum of long-range couplings over the minimum nearest neighbor couplings is approximately 8.64%, thus they can be ignored. The mode transport and confinement in SD are further simulated using the continuous model [39].

To experimentally demonstrate the aforementioned system design in SD, we apply the cw-laser-writing technique developed earlier for inducing photonic lattices laterally from the top of nonlinear crystal strontium barium niobate (SBN:61) [29]. The wiggling frequency and amplitude of the writing beam are tuned by a spatial light modulator (SLM). To clearly see the difference of mode evolution in the two different mode arrays, a long propagation distance is needed. Due to the restriction of the crystal length ($L = 20 mm$) in the experiment, the whole waveguide array is divided into several sections along the propagation direction. Each section is laterally written into the crystal in sequence, thus effectively lengthening the propagation distance (Fig. 3a). Meanwhile, we develop a cascade-probing method to illustrate the mode coupling in SD: The amplitude and phase of the output beam from one waveguide section are recorded by a camera, digitally duplicated by the SLM, and then launched into the next section of the waveguide arrays as an input (Fig. 3a). Since the duplicated beam carries all the amplitude and phase information from one section to another, such cascade probing equivalently "glues" all parts of the waveguide arrays together, therefore allowing us to examine mode evolution through long arrays [39].

The coupling strength in SD is controlled by the distance between the nearest waveguides and a proper wiggling frequency of the writing beam $\Omega = 58.70 m^{-1}$. The selected eigenmode of the lattice is generated by the SLM in the experiment, and is used as the probe beam. With the cascade probing method, the output intensity and phase distributions of the probe beam at $z = 20mm$ and $40mm$ in real space are extracted and plotted in Fig. 3(b1, b2) together with results of numerical simulations from the continuous model [39]. The deviation in the phase distributions at some sites has very minor effect on the output mode distributions since the intensity at those sites is negligible. The complex amplitude of the output beam $\psi(z) = \sqrt{I(z)} \exp(i\alpha(z))$, where $I(z)$ and $\alpha(z)$ are the measured intensity and phase, are projected onto the eigenmodes,

$$\eta_i(z) = \langle \varphi_i | \psi(z) \rangle \tag{4}$$

The overlap values $|\eta_i(z)|^2$ on every mode at $z = 20mm$ and $40mm$ are plotted in Fig. 3b3. The main components reside on the initially excited mode at $z = 20$mm. However, at a longer propagation distance ($z = 40mm$), the probe beam spreads into the neighboring modes (Fig. 3b3), in the mode array with equal spacing. The experimental results and simulations agree well.

On the other hand, in the mode array with outlying edges, the wiggling frequency of the waveguide is set to $\Omega = 30.55 m^{-1}$ (Fig. 2b2) [39]. The total propagation length is set to $L = 40mm$ in mode array of equal spacing and $L = 80mm$ in mode array with outlying edges to roughly have the same $\Omega L$ as for the array with equal spacing. Under the same initial excitation, the intensity and phase distributions at $z = 40mm$ and $80mm$ are plotted in Fig. 3c1 and c2, which again agree well with the simulations. Most importantly, the dynamics in SD (Fig. 3c3) indicates the confinement of the initially excited mode in the mode array with outlying edges, which corresponds to effective guidance of a complex beam profile observed in the experiment (Fig. 3c1, c2).

To illustrate the power of our ANN designed dynamics in SD, we modify the one-dimensional SSH lattice by ANNs to obtain bands with linear dependence of energy on the mode number and use these structures to obtain morphing of the initially excited bulk mode into a topological edge mode residing in the gap (see Fig 4.). In Fig. 4a1 and b1 we plot the eigenvalues of a trivial and nontrivial SSH lattice, respectively, with $N = 12$ sites; note that the eigenvalue separations between the modes in the bands are identical, i.e., we have linear dependence of $\beta_n$ on the mode number $n$ [39]. The nontrivial lattice (Fig. 4b1) in addition possesses two topological modes in the gap; they are topological because the conventional SSH lattice can be continuously deformed in our ANN designed lattice without closing the gap [39]. First, we excite the mode in the trivial SSH lattice, which is adjacent to the gap (Fig. 4a2) and wiggle the lattice with frequency identical to the eigenvalue spacing between adjacent modes in the bands ($\Delta\beta = \Omega$). Due to the presence of the gap and the chosen wiggling frequency, the initially excited mode can couple only to adjacent mode in its own band (it cannot cross the gap). Thus, dynamics in SD corresponds to discrete transport of light restricted to one band (Fig. 4a2). The corresponding dynamics in real space is shown in Fig. 4a3. Next, in order to couple light into a topological mode, we need to adjust the wiggling frequency as illustrated in the following example. We excite the mode in the top band of the non-trivial ANN designed SSH lattice, with the largest eigenvalue, and wiggle the lattice at the frequency that enables discrete transport in the upper band in SD (Fig. 4b2). When the mode adjacent to the gap becomes sufficiently populated, we switch the wiggling frequency to half of the gap width ($\Omega'_2 = \Delta\beta'_2$), which enables light to couple into the topological mode (Fig. 4b2). After that, we turn off the wiggling, and light remains propagating in this mode. In Fig. 4b3 we plot the corresponding dynamics in real space, which can be interpreted as morphing of light from a bulk mode into a topological edge mode merely via properly tailored control of the system (wiggling frequency).

In summary, we have theoretically proposed and experimentally demonstrated a scheme to design dynamics in an SD by employing deep learning ANNs. The scheme is based on constructing a lattice in real space that has a predesigned spectrum of mode eigenvalues. By employing judiciously chosen time-dependent perturbations, one can in principle controllably couple the modes and tailor the dynamics in engineered SDs. This is here demonstrated experimentally and theoretically in photonic lattices where an appropriate $z$-dependent wiggling tailors dynamics in SDs. More specifically, we have demonstrated SD discrete light transport, light confinement by a non-topological outlying edge mode, and finally morphing of light into a topologically protected edge mode in an ANN-designed SSH photonic lattice. The scheme based on deep learning for the construction of SDs may be extended to other parametrical spaces, which may provide insight into the study of fundamental physics that are inaccessible or even nonexistent in geometrically limited configuration spaces [20, 41]. From the viewpoint of applications, mode control in SDs, such as morphing from bulk modes into topological edge modes, has great potential for achieving mode collection, mode lasing, optical switching, and mode-division-multiplexing based data transmissions [25, 26, 42-44]. Since only the distance and frequency of curved waveguides needs to be tuned in photonic lattices, our method and results are potentially relevant for the design and fabrication of integrated photonic devices, especially with the recipes from ANN-empowered topological and synthetic dimension photonics, promising for the next decade of photonic material and device applications [35-37, 45, 46].

**Acknowledgement:**

This work was supported by National Key R&D Program of China (2022YFA1404800), the National Natural Science Foundation of China (12134006, 12274242, 11922408, 12204252), China Postdoctoral Science Foundation (BX2021134, 2021M701790), and the Natural Science Foundation of Tianjin for Distinguished Young Scholars (Grant No. 21JCJQJC00050), PCSIRT (IRT_13R29), 111 Project (No. B23045) in China.

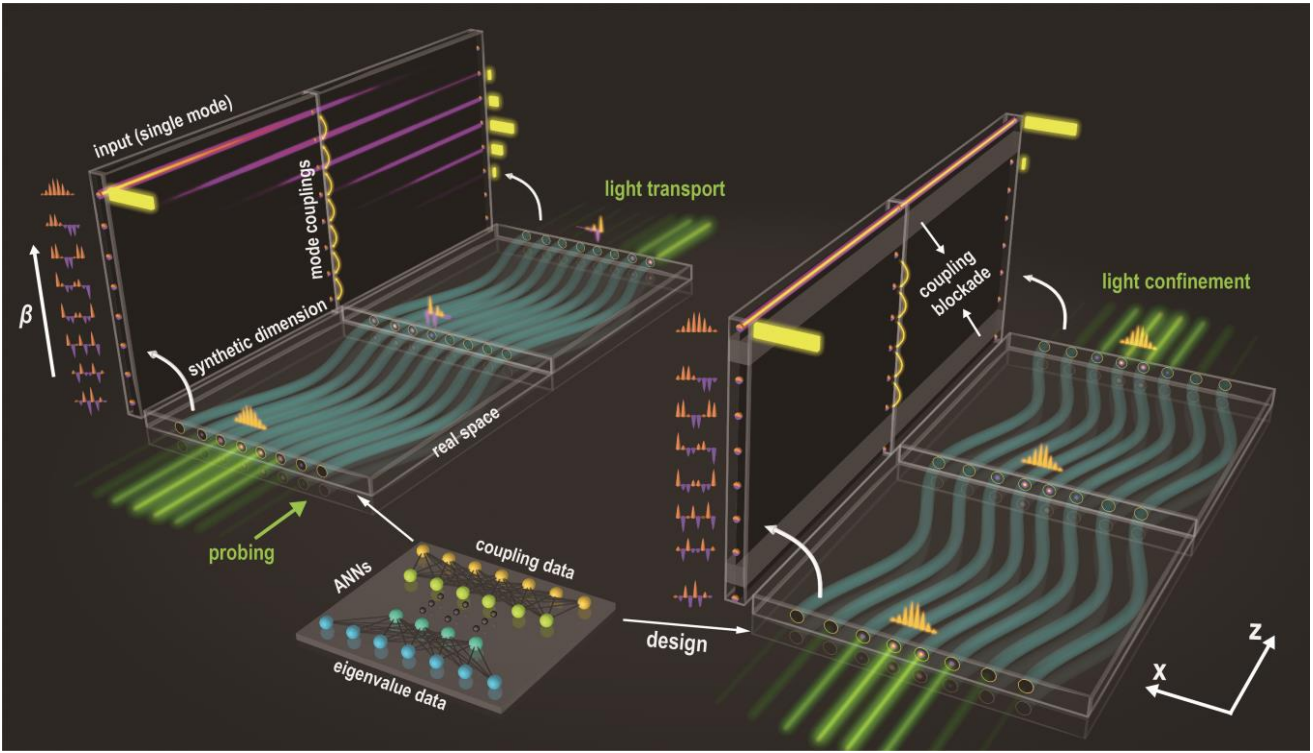

**Fig. 1 Scheme for mode manipulation in SD assisted with ANNs**. A probe beam at input representing one mode (edge in SD) is launched into different synthetic mode arrays, through which light is either transported (left) or confined (right) in SD, depending on the ANN design of the arrays. The input and output data of the ANNs are based on the preassigned eigenvalues and couplings of the arrays. Waveguides are curved along the *z*-direction in real space. Vertical planes show the mode evolution in SD, where orange/purple profiles lined up vertically are the eigenmode distributions, forming the lattices in SD. The yellow bars denote the mode distribution of the probe beam, and the shaded zone represents a coupling blockade in SD. A complex beam profile is well maintained during propagation due to the proper design of coupling blockade in the right panel.

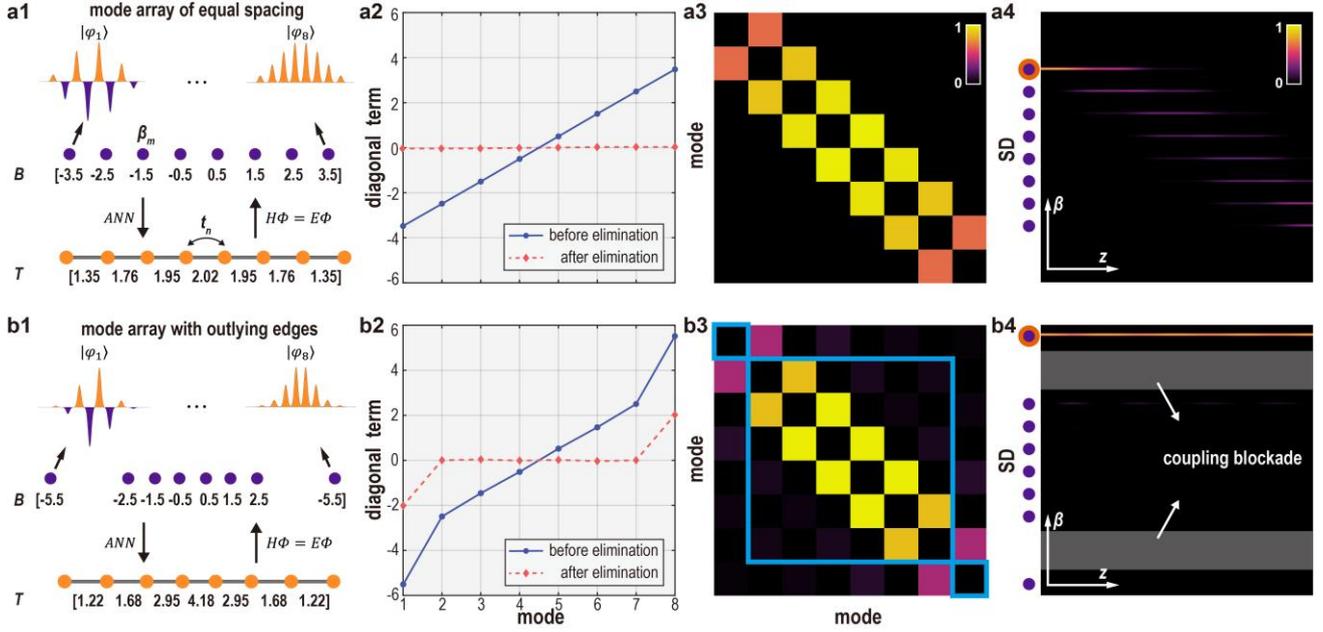

**Fig. 2 Illustration of mode evolution in different mode arrays designed by ANNs.** (a1-a4) Illustration of mode arrays with equal spacing of eigenvalues $\beta_m$. (a1) The sketch of the eigenvalue array $B$ and corresponding eigenmodes $|\varphi_i\rangle$. The arrangement of the coupling array in real space (labeled $T$) is calculated by ANNs. (a2) Distribution of $\mathrm{diag}(\Phi_A^\dagger H_0 \Phi_A)$ before (solid blue lines) and after (dashed red lines) elimination of the potential difference between eigenmodes by using a proper rotating wave frame. (a3) The off-diagonal values of $\Phi_A^\dagger H_1(z) \Phi_A$ indicate the couplings between eigenmodes in SD; we plot the normalized absolute values of the complex matrix elements. (a4) Mode evolution in SD. The orange circle indicates the initially excited mode. (b1-b4) have the same layout as (a1-a4), except that they are for the mode arrays with outlying edges, showing that the excited mode is well confined in SD. The modes in the squared region indicated in (b3) are mutually coupled, and the shaded zones in (b4) show the coupling blockade between the edge and bulk modes in SD. (b4).

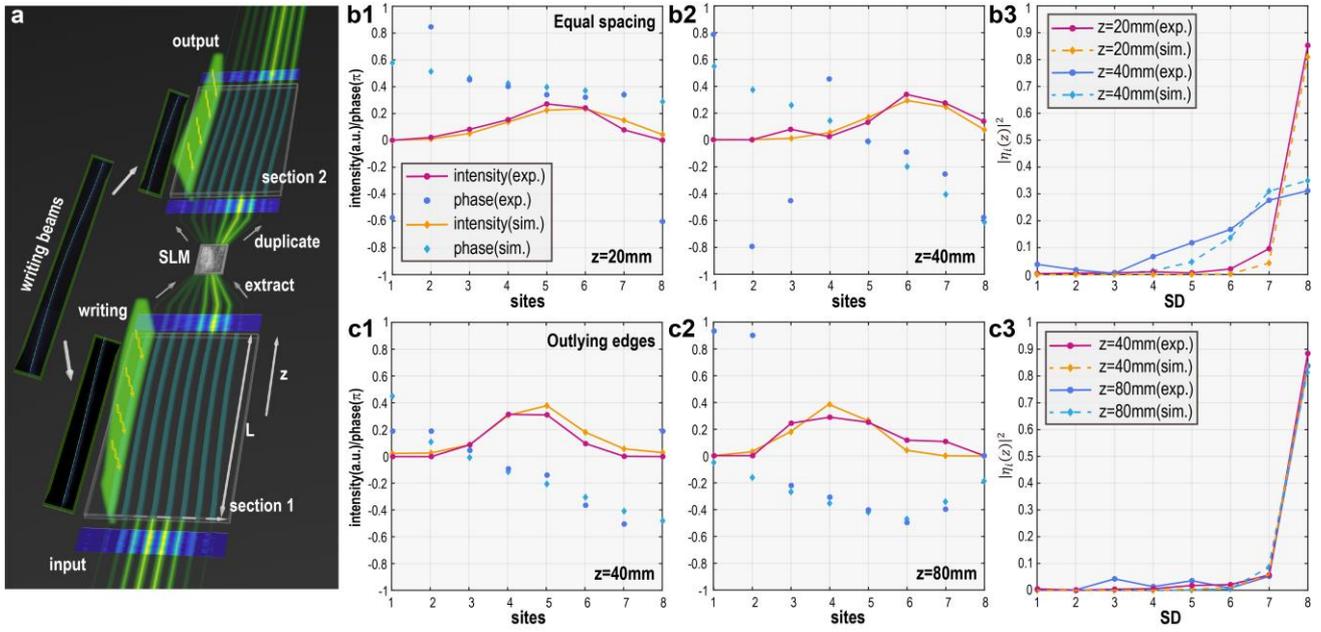

**Fig. 3 Experimental demonstration of mode manipulation in SD and corresponding simulations.** (a) Illustration of the cw-laser writing and cascade probing method in the experiment. Curved waveguide arrays are written section by section (from the top of a nonlinear crystal), and the output of the probe beam (propagating through the arrays along *z*-direction) from one section is taken as the input for the subsequent section assisted with the SLM, thus effectively increasing the propagation distance. (b) Results from the mode array with equal spacing, where (b1, b2) show the output amplitude and phase distribution from the experiment and simulation at $z = 20$mm and $40$mm. (b3) shows the corresponding output distribution in SD. (c1-c3) Results from the mode array with outlying edges, with the same layout as (b1-b3) but at even longer propagation distances, showing confinement of the excited mode in both real space and SD.

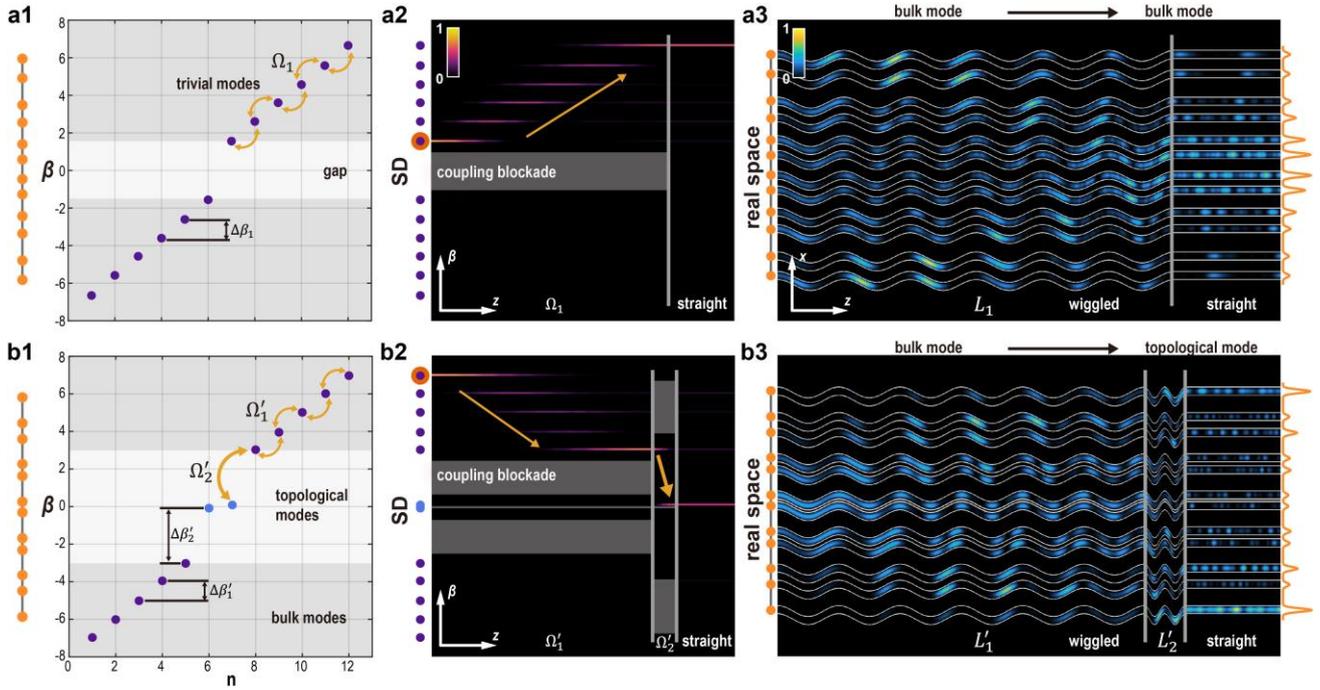

**Fig. 4: Mode switching and morphing into topological modes by tuning the array in SD.** (a) Mode switching between bulk modes in a topologically trivial lattice designed by ANNs. (a1) The lattice illustration in real space (far-left column) and corresponding eigenvalue distribution (right panel). Bulk modes above or below the gap couple to each other without a coupling blockade under an array wiggling frequency $\Omega_1 = 1$ and the eigenvalue difference $\Delta\beta_1 = 1$. (a2) Mode evolution during propagation in SD, where the orange circle indicates the initially excited mode. The second region distinguished by the vertical lines is straight waveguides. The shaded zones indicate the coupling blockades in SDs in different regions. (a3) The light evolves in real space, where $L_1 = 36.3$ is the propagation length in first region. The plot on the right shows the average intensity distribution in the straight waveguide region. (b) Same layout as (a), but in a topologically nontrivial lattice showing the morphing of bulk modes into a zero-energy topological mode where eigenvalue difference $\Delta\beta'_1 = \Delta\beta_1$, $\Delta\beta'_2 = 3\Delta\beta_1$ and the wiggling frequency $\Omega'_1 = \Omega_1$, $\Omega'_2 = 3\Omega_1$. The propagation length $L'_1 = 38.3$ and $L'_2 = 4.22$ in first and second regions in (b3), respectively.